# Software Fairness Debt

Ronnie de Souza Santos
ronnie.desouzasantos@ucalgary.ca
University of Calgary
Calgary, Alberta, Canada

Felipe Fronchetti
fronchettl@vcu.edu
Virginia Commonwealth University
Richmond, Virginia, USA

Sávio Freire
savio.freire@ifce.edu.br
Instituto Federal do Ceará
Morada Nova, Ceará, Brazil

Rodrigo Spinola
spinolaro@vcu.edu
Virginia Commonwealth University
Richmond, Virginia, USA

## ABSTRACT
As software systems continue to play a significant role in modern society, ensuring their fairness has become a critical concern in software engineering. Motivated by this scenario, this paper focused on exploring the multifaceted nature of bias in software systems, aiming to provide a comprehensive understanding of its origins, manifestations, and impacts. Through a scoping study, we identified the primary causes of fairness deficiency in software development and highlighted their adverse effects on individuals and communities, including instances of discrimination and the perpetuation of inequalities. Our investigation culminated in the introduction of the concept of software fairness debt, which complements the notions of technical and social debt, encapsulating the accumulation of biases in software engineering practices while emphasizing the societal ramifications of bias embedded within software systems. Our study contributes to a deeper understanding of fairness in software engineering and paves the way for the development of more equitable and socially responsible software systems.

## CCS CONCEPTS
• **Social and professional topics**;

## 1 INTRODUCTION
In an era where technology permeates every aspect of our lives, the expectation that software is effective in representing the multifaceted characteristics of our society has transcended technical needs and now stands as an ethical obligation for developing algorithms and systems that are both equitable and inclusive [2]. In this context, the concept of software fairness emerges as a crucial non-functional requirement and a quality attribute for software, especially those based on data-driven processes [18].

Software fairness refers to the ethical principle and practice of ensuring that software systems, algorithms, and their outcomes are just, equitable, and unbiased across different groups of people, regardless of their characteristics such as race, gender, ethnicity, or socioeconomic status [18]. In software engineering, fairness typically involves preventing discrimination, promoting inclusivity, and mitigating potential biases in the design, development, deployment, and usage of software applications and systems [11, 44]. Though not entirely new to software development, this concept has recently gained significant traction, fueled by the escalating discussions surrounding software engineering for artificial intelligence and ethics in machine learning [1, 3]—a scenario that underscores its essential role in understanding the impact of biased software in modern software engineering practices.

Software fairness has its roots embedded in the biases that can permeate throughout the stages of software development [1, 3, 11]. If left unaddressed, these biases can culminate in issues that pose significant risks and potential harm to individuals within our society [22]. Examples include discriminatory practices in criminal justice, healthcare, and financial algorithms targeting Black communities [17, 27, 29], distortions in skin tones and identifications in digital filters [36], biased hiring practices with reinforcement of gender stereotypes [43], and content moderation targeting LGBTQIA+ individuals [35], among other forms of discrimination.

In this study, we investigate the intrinsic nature of software fairness, which is positioned at the intersection of technical and social aspects within software engineering. Our inquiry focuses on identifying the underlying causes of biases that could lead to discriminatory outcomes for software users. We aim to explore the relationships among these elements, aligning our findings with established definitions of technical debt [15] and social debt [40] to foster the improvement of software development practices in this context. In particular, we are interested in answering the following research question: ***How can the concepts of technical debt and social debt be integrated into the context of software fairness?***

The subsequent sections of this paper are outlined as follows. In Section 2, we present relevant concepts that shape the scope of this research. Following this, in Section 3, we present our methodology. In Section 4, we present our main findings. In Section 5, we present the key insights derived from our analysis. Finally, Section 6 outlines our main contributions and directions for future studies.

## 2 BACKGROUND

This section introduces key concepts essential for our investigation. We start by discussing technical debt, a well-established concept in software engineering. Then, we introduce the metaphor of social debt, illustrating the broader societal implications of technical decisions. Lastly, we emphasize the possible limitations of using these debt concepts alone to address the emerging societal concern regarding software fairness.

### 2.1 Technical Debt

Technical debt is the phenomenon that occurs when software developers opt for technical shortcuts or compromises to achieve short-term gains, such as faster feature delivery or reduced costs. However, these compromises potentially compromise the long-term health of the project [7, 23]. The need for these shortcuts arises as software engineers balance competing constraints of budget, release deadlines, and software quality. Often, these shortcuts create problems with the internal quality of the software, i.e., problems that are invisible to the user but impact long-term maintenance costs and the ability to meet budget, release, and quality targets [6].

These internal quality problems are referred to as "debt" because, just like financial debt, technical debt often has to be repaid with interest in the form of higher costs and effort to maintain the software [23]. Technical debt permeates the software industry in organizations and projects of all sizes. Most software developers regularly face the consequences of technical debt [32, 34]. Recent studies demonstrated that 25% of development effort is spent—wasted—on paying interest on technical debt [9, 10].

Technical debt is generally associated with various negative impacts, including risks associated with unexpected delays in software evolution and difficulty in achieving a desired level of quality [32]. However, if teams are aware of its presence and risks [26] and are able to manage it well, technical debt can be a good investment and reduce the time or cost required to achieve project goals [21]. Conversely, unaddressed technical debt can lead to financial and technical complications, resulting in increased need for maintenance and evolution costs, ultimately jeopardizing the success of the project [9, 28].

While several types of technical debt exist, researchers have most commonly studied code debt and design debt [33]. While *code debt* describes problems that can reduce the legibility of source code and increase maintenance difficulty, e.g., code duplication, unnecessary complexity, or code that is difficult to read, *design debt* describes violations of good object-oriented design principles, e.g., complex classes or methods [4].

### 2.2 Social Debt

Social debt is a metaphor that explores the repercussions of both technical and social decisions within work environments and among individuals [41]. This phenomenon often originates from one or more suboptimal socio-technical decisions, where the accumulation of such decisions correlates with increasingly adverse impacts on software development teams. The persistence of these poor decisions exacerbates their effects over time, amplifying their intensity.

In this context, community smells denote specific relationships between these deficient socio-technical decisions and their resulting negative outcomes. As a result, the presence of these community smells contributes to the accumulation of social debt. The literature documents 30 recurring community smells [30, 38, 39, 41, 42] and represents them using a cause-and-effect model. The causes stem from adverse social and organizational situations arising from poor socio-technical decisions. Meanwhile, the effects portray the repercussions of these poor socio-technical decisions on software development teams, such as strained social interactions, misconduct, and faulty artifacts.

One example of community smell is architecture by osmosis. This smell characterizes individuals making architectural decisions without proper information management. While some team members handle and filter information via undefined communication channels or methods, others employ varying standards to document change requests. These divergent approaches lead to difficulties in identifying the sources of architecture decisions, waste of time, and unstable architecture configurations [38].

### 2.3 Ramifications of Technical and Social Debt

While technical debt and social debt can indirectly impact software users by leading to reduced software quality and a poor user experience, their primary effects are predominantly felt at the software development level. Technical debt affects the codebase and software design [4, 33], whereas social debt influences team performance and interactions within the development environment [38]. However, the emergence of AI-powered software solutions introduces a shift where shortcuts in development may *directly* impact users, potentially leading to increased costs and debt items that need to be repaid promptly.

## 3 METHOD

This study is a scoping review [31] designed to identify and explore previous research published over the years that investigated software fairness and explored its roots, examples, and effects. This type of review is used to consolidate findings from various primary studies, including experiments, case studies, surveys, and experience reports, among others [25, 37]. In this study, we adhered to the established guidelines for systematic reviews in software engineering [25], as detailed below.

### 3.1 Specific Research Questions

We are building this scoping study to offer a comprehensive overview of research on software fairness, focusing on studies published in leading software engineering conferences and journals. Hence, to guide our data collection, data analysis, and synthesis of evidence, we developed five specific research questions:

- *RQ1*: How has the number of studies on software fairness changed over time in software engineering?
- *RQ2*: What are the various definitions of fairness documented in the software engineering literature?
- *RQ3*: What are the underlying causes for the absence of software fairness in software systems?

- *RQ4*: What are the effects of software fairness on the outcomes of software systems?
- *RQ5*: What examples of bias and discrimination have been reported in the context of software fairness in software engineering?

Later, these questions supported us in interpreting the insights derived from the literature findings to address our overarching research question, i.e., to establish the relationship between software fairness and technical and social debt based on the evidence presented in the literature.

### 3.2 Data Sources and Search Strategy

Our search for studies on software fairness entailed a manual process focused on major software engineering scientific forums. Specifically, we targeted conferences such as ICSE, FSE, ESEM, and ICSME, along with journals including EMSE, TSE, and IEEE Software. Employing a manual approach enabled us to conduct a targeted review of the literature, aligning with the primary channels utilized by the software engineering community. In this process, we manually accessed the proceedings of the above-mentioned conferences and journals, covering one decade of publications (e.g., the years from 2014 to 2023). The search was conducted based on the metadata of the studies, including title, abstract, and keywords, and it was concluded in early 2024.

### 3.3 Inclusion and Exclusion Criteria

We applied three exclusion criteria and one inclusion criterion to evaluate the papers identified through our search. First, papers were excluded from our research if they met any of the following exclusion criteria: a) *EC-1* Unavailable for download or inaccessible for online reading; b) *EC-2* Incomplete texts, like drafts, presentation slides, or abstract only; c) *EC-3* Proposing tools for a particular scenario without offering a thorough discussion of the problem.

Following the application of exclusion criteria, our focus was specifically on selecting papers that tackled issues related to software fairness, such as biases and algorithmic discrimination. Additionally, we included papers that offered insights, lessons learned, or experience reports on the theme. Some papers were excluded during the exclusion/inclusion criteria phase as they did not present substantial evidence on the investigated topic. Instead, they merely referenced the topic or provided vague information without exploring the implications of software fairness.

### 3.4 Data Extraction

After completing the selection phase, we introduced a structured form to facilitate the data extraction process. At this stage, we accessed the full text of each paper and documented relevant information for our study. Four researchers collaborated on this task to ensure the accuracy of data extraction, thereby enhancing the reliability of our results. Initially, three researchers worked independently and extracted data from one-third of the identified studies. Subsequently, a fourth researcher extracted data from a randomly selected sample of papers obtained from the entire pool of studies. A comparison between both extractions was then conducted to assess alignment and verify reliability. Any discrepancies in the extraction were resolved through comprehensive discussion and consensus. Overall, only a few discrepancies emerged during the data extraction process involving two papers. These disagreements were resolved with the participation of a third researcher.

### 3.5 Data Analysis

Upon completing the form with the evidence gathered from the papers, we employed two data analysis approaches to synthesize the obtained data and present our findings. We began by applying descriptive statistics [19] to summarize the distribution and frequency of papers by year, publication venue and the domain of software engineering focused on in the paper. This analysis provided insights into the data by partitioning papers into sub-groups using various statistical functions, including means, proportions, totals, and ratios. Additionally, it allowed us to establish connections among the categories identified during the qualitative analysis.

Following this, we applied qualitative analysis to the data extracted from the papers using coding strategies [20] and thematic analysis [14]. Initially, we employed line-by-line coding [20] to thoroughly explore the data and construct codes reflecting specific details about software fairness, such as definitions, causes, effects, and examples. Subsequently, we utilized thematic analysis [14] to delve deeper into the codes, drawing cross-references and comparisons among them. This process enabled us to integrate the codes into well-defined themes or categories.

## 4 FINDINGS

We identified 39 studies focusing on software fairness, published between 2017 and 2023. During this timeframe, there has been a noticeable growth in interest in the topic, starting with only one paper published in 2017 and peaking with 13 papers published in 2022. Among the venues, 17 studies were published at ICSE, 12 at FSE, and one at ICSME. Additionally, regarding the journals, four articles were published in EMSE, four in TSE, and one in IEEE Software. The majority of the identified studies were published in the context of software testing (22 papers), followed by software design and implementation (10 papers), and software engineering in general (7 papers). Table 1 summarizes the main findings of our mapping. Some direct references to the analyzed studies are not directly mentioned here due to template and page restrictions, however, our dataset is available at: https://figshare.com/s/ebf820a8d85a601040c2

### 4.1 Software Fairness Definitions

In software engineering, studies often center on two core aspects for defining fairness. On the one hand, we identified 29 papers that elaborate on the idea of fairness using the concept of individual and group fairness as guiding principles. On the other hand, we found eight studies that take an implementation perspective, leveraging general elements of software design to formulate fairness criteria. Two studies did not present a structured definition for the term.

The definition of fairness, rooted in the concepts of individual and group fairness, underscores the fundamental premise that systems should abstain from discriminating between different individuals or groups based on protected attributes. This means that characteristics such as race or gender should not influence software outcomes. Therefore, individual fairness requests that two individuals who have similar characteristics or features that are relevant to the decision-making process may obtain similar outcomes, while group fairness aims for equitable treatment across groups divided by protected attributes. In other words, the distribution of outcomes produced by software should be consistent across different demographic groups.

**Table 1: Mapping Summary**

| Finding | Theme | Papers |
| --- | --- | --- |
| Definitions | Individual-Group Based Concept | FAIR02 FAIR05 FAIR06 FAIR07 FAIR09 FAIR11 FAIR12 FAIR13 FAIR14 FAIR15 FAIR16 FAIR17 FAIR18 FAIR20 FAIR21 FAIR23 FAIR24 FAIR25 FAIR26 FAIR27 FAIR28 FAIR29 FAIR30 FAIR31 FAIR32 FAIR36 FAIR37 FAIR38 FAIR39 |
| | Implementation Perspective Concept | FAIR01 FAIR03 FAIR04 FAIR08 FAIR10 FAIR19 FAIR22 FAIR35 |
| Root Causes | Cognitive Bias | FAIR01 FAIR02 FAIR13 FAIR25 FAIR37 |
| | Design Bias | FAIR01 FAIR04 FAIR06 FAIR09 FAIR10 FAIR11 FAIR12 FAIR13 FAIR18 FAIR20 FAIR22 FAIR23 FAIR29 FAIR39 |
| | Historical Bias | FAIR01 FAIR17 FAIR26 |
| | Model Bias | FAIR05 FAIR11 FAIR14 FAIR15 FAIR19 FAIR21 FAIR30 FAIR35 FAIR38 |
| | Requirement Bias | FAIR01 FAIR12 FAIR31 |
| | Societal Bias | FAIR13 FAIR36 |
| | Testing Bias | FAIR10 FAIR11 FAIR24 |
| | Training Bias | FAIR03 FAIR07 FAIR08 FAIR09 FAIR12 FAIR13 FAIR28 FAIR32 FAIR34 FAIR36 |
| Effect | Exacerbation of Social Inequality | FAIR16 FAIR28 FAIR35 FAIR36 FAIR38 FAIR39 |
| | Legal Concerns | FAIR26 FAIR30 FAIR32 FAIR35 FAIR39 |
| | Limited Algorithmic Reliability | FAIR17 FAIR20 FAIR28 FAIR30 FAIR31 FAIR32 FAIR35 FAIR37 |
| | Proliferation of Discrimination | FAIR01 FAIR02 FAIR03 FAIR04 FAIR05 FAIR06 FAIR07 FAIR09 FAIR10 FAIR11 FAIR12 FAIR13 FAIR14 FAIR21 |
| | Psychological Harms | FAIR35 |
| | Reduced Algorithmic Accuracy | FAIR12 FAIR15 FAIR19 FAIR20 FAIR25 FAIR28 FAIR30 FAIR31 FAIR32 FAIR34 FAIR35 FAIR39 |
| | Reinforcement of Stereotype | FAIR01 FAIR08 FAIR09 FAIR17 FAIR19 FAIR26 FAIR29 FAIR30 FAIR31 FAIR33 FAIR34 FAIR35 FAIR36 FAIR37 FAIR38 FAIR39 |
| Examples | Ageism | FAIR25 |
| | Classism | FAIR02 FAIR05 FAIR10 FAIR12 FAIR13 FAIR33 |
| | Racism | FAIR01 FAIR03 FAIR04 FAIR07 FAIR10 FAIR12 FAIR12 FAIR13 FAIR15 FAIR17 FAIR19 FAIR20 FAIR29 FAIR30 FAIR37 |
| | Sexism | FAIR01 FAIR03 FAIR12 FAIR13 FAIR15 FAIR17 FAIR18 FAIR19 FAIR23 FAIR27 FAIR28 FAIR34 |
| | Xenophobia | FAIR12 FAIR24 |

The definition of software fairness, grounded in specific aspects of software design and implementation, is tied to the characteristics of models or metrics utilized during system construction. This entails addressing challenges stemming from misled training algorithms, which may rely on one or more potential bias features. Within this concept, fairness is understood as a collection of various strategies and techniques employed to foster the development of more equitable and inclusive data-driven systems. Additionally, fairness is often evaluated with respect to each output characteristic individually, emphasizing the need to identify a set of protected attributes to define fairness accurately.

Our analysis revealed that diverse definitions of fairness can lead to different outcomes in the proposed systems and techniques for addressing the problem in the software engineering literature. Some studies are deeply aligned with the social aspects of the fairness problem, while others predominantly focus on technical considerations. This diversity of focus underscores the multidimensional nature of fairness concerns in software engineering and suggests that the field is grappling with balancing technical aspects with ethical and societal considerations.

### 4.2 Root Causes of Software Fairness Deficiency

In software development, the imbalance of fairness can stem from various origins, spanning across different facets of the development process. These sources may not only originate within the software environment itself but can also be influenced by external factors. Lack of fairness can be rooted in various aspects, including the software practices, the technologies utilized, and the dynamics among team members. Below, we describe the root causes identified in our analysis:

- *Cognitive Bias*: Developers may hold preconceived judgments or attitudes that sway their decisions in algorithm construction, potentially leading to the perpetuation of inequalities. Furthermore, these biases may impede their ability to identify and address issues that contribute to disparities within software systems, further exacerbating inequalities in their implementation and impact.
- *Design Bias*: Preconceived notions or technical preferences can significantly sway architectural decisions and user interface designs in software systems, inadvertently reinforcing inequalities. For instance, messages in the interface may cater to specific demographics, while values may not accommodate the diversity within user groups or even features prioritizing certain demographics over others.
- *Historical Bias*: Previous events or data experiences have the potential to influence decisions and outcomes within software systems, potentially perpetuating discrimination. This bias can manifest in various ways, such as favoring certain user groups based on historical data patterns or perpetuating discriminatory practices embedded in legacy systems or past software development decisions.
- *Model Bias*: Systematic errors or inaccuracies in mathematical or computational models used in software applications can lead to skewed predictions or analyses, potentially perpetuating inequalities. These biases may arise from inherent limitations in the design or implementation of such models, resulting in biased outcomes that disproportionately impact certain user groups or demographics.
- *Requirements Bias*: Inherent bias in project specifications or user requirements might potentially impact the definition or prioritization of functionalities, resulting in perpetuating discrimination. Such bias may arise from limited diversity or lack of inclusiveness in the requirements-gathering processes, leading to the development of software systems that fail to adequately treat groups of users fairly.
- *Societal Bias*: Discriminatory practices embedded within social structures or norms shape the software context and its outcomes, potentially perpetuating inequalities. For instance, biases in hiring practices may lead to underrepresentation of certain demographic groups in software development teams, resulting in products that do not adequately address the needs of those communities. Additionally, algorithms trained on biased datasets may perpetuate stereotypes or discriminate against marginalized groups, reinforcing existing societal inequalities.
- *Testing Bias*: Unfair behaviors that go undetected during the testing process due to the absence of fairness-focused testing strategies potentially perpetuate inequalities. For example, if testing criteria fail to account for diverse user demographics or scenarios, the software may inadvertently favor certain groups, reinforcing existing disparities.
- *Training Bias*: Inaccuracies within datasets used to train algorithms result in discriminatory outcomes in software applications and reinforce inequalities. For example, if training data predominantly represents one demographic group, fails to represent certain individuals in the population correctly, or is biased from its source, the resulting algorithm will likely disadvantage individuals in our society.

Addressing these biases is essential to prevent discriminatory effects in our society, which can range from minor inconveniences to life-threatening scenarios, including exacerbating social divisions and harming vulnerable populations, as our findings demonstrate in the next section.

### 4.3 Effects of Software Fairness

By implementing fairness-focused strategies to address the root causes of biases in software development, we can begin mitigating several risks and create software systems that serve all members of society fairly and equitably. In our analysis, we identified the following effects arising from the lack of fairness:

- *Exacerbation of Social Inequality*: Neglecting software fairness exacerbates existing disparities by favoring certain groups or perpetuating historical inequalities, deepening social divisions and inequities.
- *Legal Concerns*: Ignoring fairness raises legal concerns regarding discrimination and privacy violations, triggering debates on regulatory frameworks and accountability in the digital age.
- *Limited Algorithmic Reliability*: Failure to address fairness undermines algorithm reliability, leading to errors and inconsistencies that diminish user trust in decision-making processes.
- *Proliferation of Discrimination*: Lack of fairness leads to the perpetuation of discrimination, resulting in unequal treatment based on individual characteristics like race or gender, which exacerbates inequalities, hindering progress toward a more equitable society.
- *Psychological Harms*: Reduced software fairness might expose users experiencing discrimination or unfair treatment to psychological distress, including frustration and anxiety, and such emotional burdens can have detrimental effects on mental well-being and reduced quality of life overall.
- *Reduced Algorithmic Accuracy*: The absence of fairness compromises algorithmic accuracy by allowing biases to influence decision-making processes, resulting in outcomes that deviate from the intended system objective and lead to significant negative consequences.
- *Reinforcement of Stereotypes*: Disregarding fairness can perpetuate preconceived assumptions about individual or group identity, undermining current societal efforts to foster diversity and inclusion. This can lead to systemic discrimination and marginalization of individuals, hindering progress towards a more equitable society.
- *Weakening of Justice*: Neglecting fairness weakens principles of justice by introducing partiality into decision-making processes, infringing on fundamental rights and freedoms, which diminishes trust in institutions relying on these systems and compromises the integrity of legal and ethical frameworks.

As outlined above, failing to address the negative impacts of biases through the implementation of software fairness can lead to severe consequences for society. Our findings demonstrate that accounting for fairness in software development is essential to mitigate these adverse effects and foster more just and equitable outcomes for all stakeholders.

## 4.4 Examples of Algorithmic Discrimination in the Software Engineering Literature

We identified examples of algorithmic discrimination in the software engineering literature, shedding light on the pervasive biases and prejudices within software systems. We focused on extracting and discussing real-world examples used in the studies to fuel discussions. For instance, studies mentioned ageism, which is characterized by bias or discrimination against individuals based on their age, particularly targeting older adults; classism, characterized by perpetuated discrimination based on social or economic class; racism, which involves biases against individuals based on race or ethnicity; sexism, which is characterized by biases against individuals based on their sex or gender, typically against women or girls; and Xenophobia, which involves biases that lead to hostility towards foreigners, such as those from different cultures or nationalities.

It is important to note that these examples are identified within the software engineering literature. In a broader context, bias can lead to several other types of discrimination, including homophobia, transphobia, and sizeism, among others [16]. These issues are discussed extensively in various fields, such as education, health, law, and other social sciences, highlighting the interdisciplinary nature of addressing biases and discrimination in society. Additionally, 15 papers did not provide explicit examples of the type of discrimination they aimed to address or concrete instances of bias effects in society, such as discussions based on real-world examples. This suggests a disconnect between the technical aspects of research on software fairness and the real societal impact of the proposed solutions.

## 5 DISCUSSIONS

In this section, we discuss our findings. We start by revisiting the literature on technical and social debt and establishing the connection between these concepts and the characteristics of software fairness, which guided us to define the concept of software fairness debt. Following this, we discuss the implications of our findings. Finally, we address the limitations of our study.

### 5.1 Unveiling Software Fairness Debt

The management of software-reliant systems, which are becoming increasingly complex, requires better ways to manage the long-term effects of short-term expedients. The concept of technical debt has become widely recognized as a useful tool for comprehending and addressing these challenges. It gained rapid acceptance within the industry, facilitating discussions between business and technical stakeholders about software quality [13]. On the other hand, even though the study of software social debt is recent [40], it has the potential to become the source of prolific human-centric research in software engineering [12].

Both technical and social debt share similarities, as they are often ignited by risky decision-making processes that impact the workforce engaged in software development tasks and the resultant software quality. Nevertheless, there are differences. Technical debt primarily centers on deferred software architecture decisions and the risk of potential software failures. In contrast, social debt is concerned with the interplay of social and technical decisions within software development environments, subsequently influencing the behavior of individuals involved.

The aforementioned characteristics of software technical and social debt are reflected in their constituents' properties: principal and interest. Although their core definition remains the same (e.g., principal - referring to the cost to eliminate the debt and interest - referring to the potential penalty for the project), the manner in which these elements are addressed in projects often diverges. Technical debt repayment typically involves activities such as code refactoring, design enhancements, and updating system documentation. Conversely, social debt repayment predominantly involves socio-technical decisions, predictive models for software productivity, and team composition management. A similar behavior is noticed for interest. Technical debt commonly results in penalties such as increased effort and reduced productivity, while social debt impacts team dynamics and contributes to the emergence of technical debt.

Therefore, while technical debt predominantly focuses on software intricacies, social debt shifts its focus to humans who participate in software development. In this context, we observed that *while technical debt and social debt provide valuable insights, they fall short of encapsulating the broader impact of software solutions on society*, which is a critical consideration in today's landscape. This observation underscores the need for a new paradigm, denoted as software fairness debt, which aims to address the broader societal implications of debt items present in software solutions. Figure 1 depicts where software fairness debt is positioned in comparison with technical and social debt. In summary, the nature of software fairness debt is:

- Causes: focused on various types of bias, representing the multifaceted origins of fairness deficiencies within software systems.
- Central concern: focused on the impact of the debt on society and its profound implications for individuals and communities.
- Effects: the actual impact resulting from the presence of this type of debt, e.g., the spectrum of outcomes ranging from minor inconveniences to enormous societal injustices.

These characteristics disjoint from those present in technical and social debt. This makes software fairness debt a different type of debt. It does not fit as a sub-type of technical or social debt. Software fairness debt can be expressed through equivalent constituent properties (e.g., principal and interest), and although it can impact software quality, its repayment would likely involve a combination of technical and social measures tailored to address biases in software systems. This could include activities such as auditing algorithms for bias, improving diversity and inclusion in software development teams, refining data collection and processing methods to mitigate bias, implementing fairness-aware machine learning techniques, training software professionals to recognize discrimination, and developing transparent and accountable decision-making processes.

Therefore, equipped with the evidence collected through our scoping study and inspired by the definition of technical debt [8], our definition of software fairness debt read as: **In software systems, fairness debt is a collection of design, implementation,** or managerial constructs that offer short-term benefits but set up a technical context that can make system changes costly or unfeasible, with significant societal impacts – an actual or contingent liability whose impact encompasses both societal and internal system qualities. Figure 2 presents a conceptual map detailing our definition of software fairness debt and its elements and relationships.

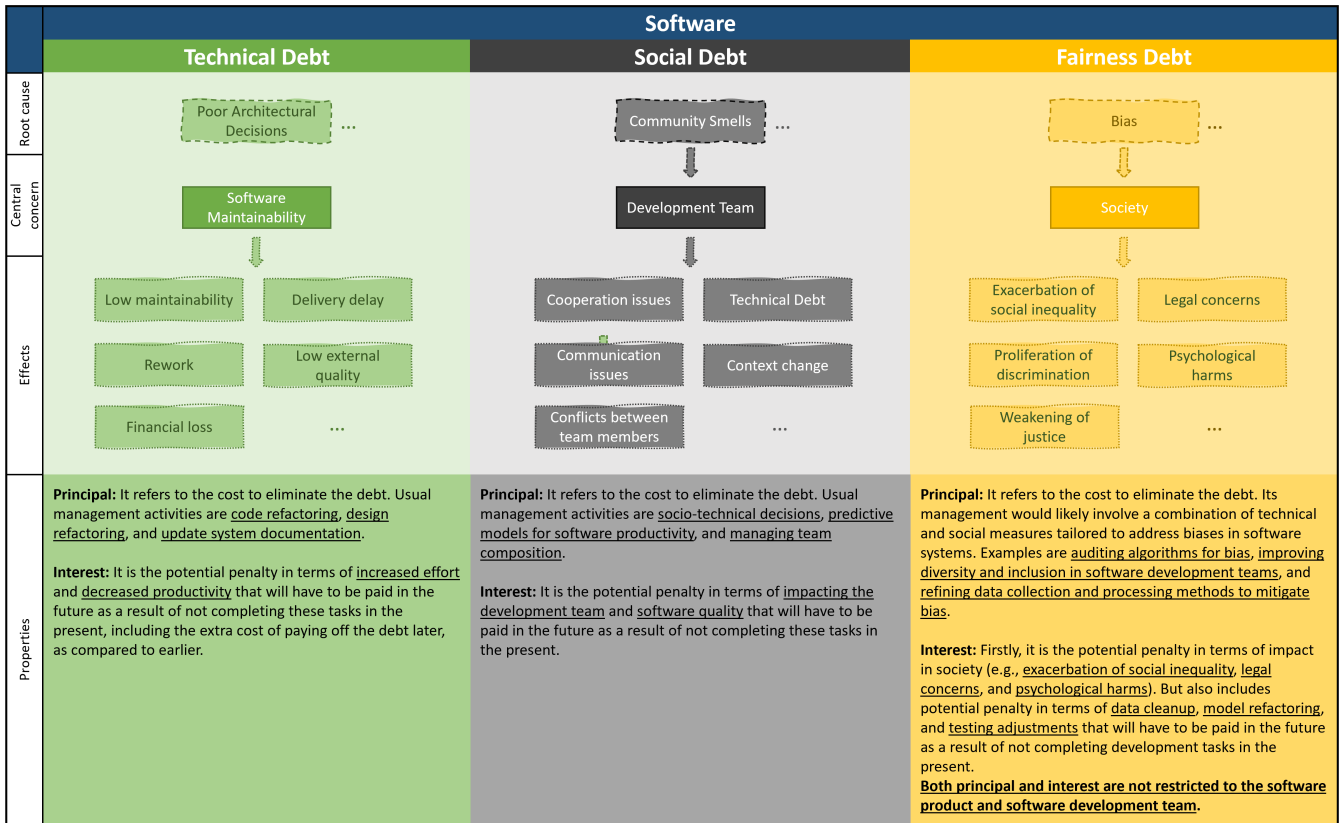

Figure 1: Software technical, social and fairness debt landscape.

This conceptual map illustrates the existing comprehension of software fairness derived from the literature reviewed in our scoping study, presented in a simplified format. The map delineates the understanding in terms of causes prompting development teams to accumulate software fairness debt (depicted in blue), the definition of fairness debt, instances of its occurrence (in yellow), and the ramifications of its presence (in orange).

Our current understanding indicates that software fairness debt can arise from various types of biases, including cognitive, design, historical, model, requirement, societal, testing, and training biases. Its manifestation can lead to numerous consequences for the project, including exacerbation of social inequality, legal concerns, and proliferation of discrimination, among others. Typical examples of software fairness debt occurrences encompass ageism, classism, racism, sexism, and xenophobia.

## 5.2 Implications

The main implication of our study is the emergence of the concept of fairness debt in software engineering, which stems from discussions about software fairness published over the years and is aligned with the transformative advancements observed in software engineering amid the rise of machine learning-based systems. In this scenario, our study sheds light on the notion of fairness debt, offering software engineering research and practice with the understanding of its causes, central concerns, and effects.

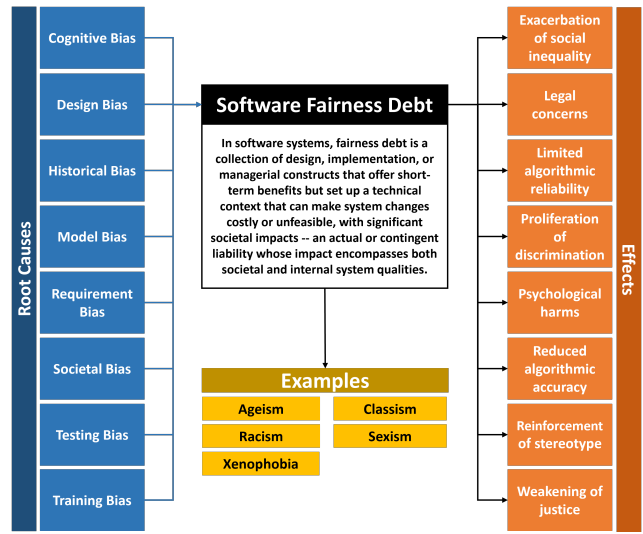

Figure 2: Conceptual map of identified elements of Software Fairness Debt.

*5.2.1 Implications to Research.* By delineating these aspects, our study presents implications for software engineering research, offering foundational insights to explore fairness debt. This includes identifying various types of bias contributing to fairness debt accumulation, understanding its central concerns, such as its impact on software quality and society as a whole, and recognizing the potential repercussions of allowing fairness debt to persist in software systems. Such insights enable researchers to develop strategies to mitigate software fairness debt accumulation, fostering the development of more equitable and socially responsible software systems.

Through the introduction of the concept of fairness debt, we are bridging the gap between the technical aspects of algorithmic bias and its societal ramifications. Our findings highlighted instances in the software engineering literature where the issue is tackled in isolation, almost detached from its real-world effects and societal implications. Hence, by integrating technical and social debts into fairness debt, we equip researchers with insights to address the multifaceted nature of bias in software systems, thus facilitating the development of more informed, ethical, and socially responsible software solutions.

*5.2.2 Implications to Practice.* This research brings significant implications for software development practice, particularly in raising awareness about the costs associated with the absence of fairness in software projects. By understanding these costs, practitioners can make more informed decisions during the software development life-cycle and allocate resources effectively to address fairness concerns early on. This proactive approach not only helps mitigate potential legal and ethical risks but also enhances the overall quality and trustworthiness of software products.

Moreover, this research is particularly important for software companies operating in today's increasingly diverse and socially conscious landscape. As modern society demands more transparency and accountability from technology providers, demonstrating a commitment to fairness becomes essential for maintaining trust and credibility. By integrating fairness considerations into their development processes and addressing fairness debt, the software industry can not only uphold ethical standards but also mitigate risks associated with biased algorithms. This approach can significantly impact various fields that rely on software systems powered by artificial intelligence. For instance, several problems have been reported recently caused by these systems in the areas of healthcare, finance, and criminal justice.

### 5.3 Limitations

Threats to validity in systematic literature reviews, including scoping studies, can include issues related to the search and selection processes, as well as the validity of the data extracted from the papers and the synthesis of evidence, including quality assessment [5]. Our main limitation lies in primarily relying on a manual selection process, targeting specific conferences and journals. However, this approach has proven successful in the literature [24], particularly when the study focuses on a very specific context like ours. By concentrating on top software engineering conferences and journals, we not only enhanced the quality of the analyzed papers but also reviewed the most relevant research on fairness in software engineering recently published. As for other threats to validity, we endeavor to minimize risks associated with data extraction and synthesis by employing a team of four researchers who cross-check papers multiple times to ensure reliability, and any discrepancies are resolved through consensus meetings.

### 5.4 Future Work and Challenges

Initially, we expect to focus our future work on a deeper exploration of the concept of software fairness debt and its implications. This includes investigating methods for categorizing, quantifying, and measuring fairness debt accumulation in software systems, as well as developing strategies for mitigating its effects throughout the software development life-cycle. Additionally, we plan to explore the economic and societal costs associated with fairness debt, providing researchers and practitioners with a comprehensive understanding of its impact on both the technical and ethical dimensions of software engineering. Moreover, from a broader perspective, our research presents opportunities for future investigations to target specific types of bias identified in our scoping study, including cognitive, design, historical, model, requirement, societal, testing, and training biases. Finally, there is a significant opportunity to explore other forms of algorithmic discrimination that remain unexplored in the context of software engineering, such as homophobia, disability discrimination, and other biases not reviewed in our findings.

## 6 CONCLUSIONS

In this work, we provide a comprehensive view of the current state of software fairness research in software engineering. Our scoping study has brought forth the main causes of bias in software development, as well as the effects of such bias in our society. Through this investigation, our main contribution is the proposition of the concept of **software fairness debt**. This concept provides researchers and practitioners with valuable perspectives for tackling the intricate dimensions of bias within software systems.

Software fairness debt complements the concepts of technical and social debt. However, it uniquely emphasizes the societal ramifications of bias embedded within software systems. Just as the metaphors of technical and social debt have sparked deeper conversations about maintainability and software team dynamics, we anticipate that the emerging metaphor of software fairness debt will catalyze the development of software solutions that are not only more informed and ethical but also more socially responsible.

In the future of software engineering, one of the key challenges for research on software fairness debt lies in developing robust methodologies to quantify and mitigate its impact effectively. Unlike technical and social debt, which have established frameworks for identification and repayment, software fairness debt poses unique challenges due to its intricate connection with societal biases and inequalities, requiring interdisciplinary collaboration among software engineers, fields being impacted by system biases, and policymakers to develop comprehensive strategies that consider both the technical and ethical dimensions of the problem.